\documentclass[aps,prl,twocolumn,showpacs,10pt,showkeys,preprintnumbers,amsmath,amssymb]{revtex4-1}

\usepackage{graphicx}
\usepackage{dcolumn}
\usepackage{bm}
\usepackage{verbatim}
\usepackage[version=3]{mhchem}
\usepackage{subfigure}
\PassOptionsToPackage{linktocpage}{hyperref}
\usepackage[hyperindex,breaklinks]{hyperref}

\begin{document}


\title{Dealloying of Platinum-Aluminum Thin Films \\ Part I. Dynamics of Pattern Formation}

 \author{Henning Galinski}
 \email{henning.galinski@mat.ethz.ch}
 \author{Thomas Ryll}
 \author{Lukas Schlagenhauf}
 \author{Felix Rechberger}
 \author{Sun Ying}
 \author{Ludwig J. Gauckler}
 \affiliation{Nonmetallic Inorganic Materials, ETH Zurich, Zurich, Switzerland}
 
 \author{Flavio C.F. Mornaghini}
 \author{Yasmina Ries}
 \author{Ralph Spolenak}
 \affiliation{Laboratory for Nanometallurgy, ETH Zurich, Zurich, Switzerland}
 
 \author{Max D\"obeli}
 \affiliation{Ion Beam Physics, ETH Zurich, Zurich, Switzerland}

\date{\today}

\begin{abstract}

Applying focused ion beam (FIB) nanotomography and Rutherford backscattering spectroscopy (RBS) to dealloyed platinum-aluminum thin films an in-depth analysis of the dominating physical mechanisms of porosity formation during the dealloying process is performed. The dynamical porosity formation due to the dissolution of the less noble aluminum in the alloy is treated as result of a reaction-diffusion system. The RBS analysis yields that the porosity formation is mainly caused by a linearly propagating diffusion front, i.e. the liquid/solid interface, with a uniform speed of $v_{\text{f}}=42(3)$~nm/s when using a 4M aqueous \ce{NaOH} solution at room temperature. The experimentally observed front evolution is captured by the normal diffusive Fisher-Kolmogorov-Petrovskii-Piskounov (FKPP) equation and can be interpreted as a branching random walk phenomenon. The etching front produces a gradual porosity with an enhanced porosity in the surface-near regions of the thin film due to prolonged exposure of the alloy to the alkaline solution.

\end{abstract}
\maketitle


In his original work, Murray Raney in 1927 dealloyed Nickel-Aluminum alloys with concentrated sodium hydroxide in order to derive extremely porous and nanostructured Nickel catalysts~\cite{Raney1}. This process has gained renewed attention in recent years for the formation of porous metallic thin films~\cite{Erlebacher1,Erlebacher2,Erlebacher3,Eilks1,Forty1,Simmonds1,Thorp1}. The interest is not only motivated by industrial needs of miniaturized sensors and catalysts but also by fundamental interests in the physical mechanisms that control the pattern or porosity formation~\cite{Erlebacher3,Eilks1}. Dealloying can be interpreted as a reaction-diffusion process, where the less noble metal in a solid solution is dissolved at the solid/liquid interface to an acid or alkaline solution. The reaction and diffusion processes can be treated in one dimension in the case of thin films. For normal diffusion, the reaction-diffusion equation (RDE) reads
\begin{equation}
	\frac{\partial}{\partial t}c(x,t)=R(c)+D\Delta_{x}c(x,t),
	\label{eq:fkpp0}
\end{equation}
where $c(x,t)$ is the local concentration, $R(c)$ is a system specific reaction term and $D$ the diffusion coefficient~\cite{Benguria1,Fisher1,Doering1}. In contrast to its importance to basic mechanisms, reliable measurements of the dealloying dynamics are rare and mostly indirect. It is usually assumed that the dealloying of the less noble metal in the alloy can be treated in terms of a phase separation at the reaction interface via a mean field approach described by the non-linear Cahn-Hilliard equation~\cite{Erlebacher1,Erlebacher3,Eilks1}. However, there is no experimental data available that conceives the dynamics of dealloying as a direct result of a reaction-diffusion (RD) system. Therefore the main objective of this paper is to prove a relation between the observable macroscopical changes in the composition and morphology of the thin film with the predictions of a reactive-diffusion equation.
The Pt/Al system has been selected, since it shows a nearly complete miscibility between the two elements including the formation of various intermetallic phases. Pt/Al layers of $300$~nm in thickness, were deposited at room temperature by magnetron co-sputtering ($P_{\text{Pt}}=37\text{~W},P_{\text{Al}}=252\text{~W},p_{\text{Ar}}=2.6\cdot10^{-3}$~mbar) onto amorphous \ce{Si3N4} substrates that were pre-cleaned using isopropanol and acetone.\\
In order to achieve a measurement scheme of sufficient significance, the substrates coated with the Pt/Al thin film were dealloyed in $4M$ \ce{NaOH} at room temperature in a time domain between $1-10$~s in steps of $\Delta t=1~$s. The reaction between the thin film and the basic solution can be established to
\begin{equation}
\ce{Pt_{x}Al_{y} ->[RT,OH-] Pt_{x+z}Al_{y-z} + $z\:$Al(OH)3}.
	\label{eq:deall1}	
\end{equation}
\begin{figure}[t!]
  \begin{center}
    \subfigure[]{\label{fig:dealloy1-a}\includegraphics[scale=1.20]{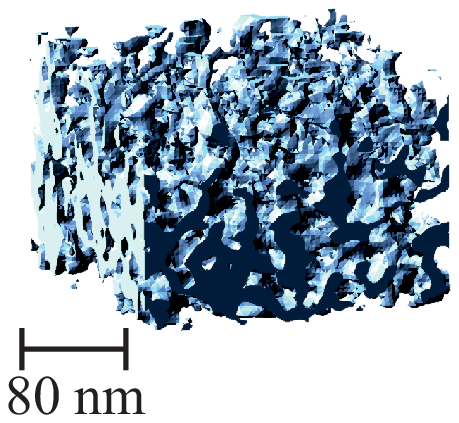}}
    \\
    \subfigure[]{\label{fig:dealloy1-b}\includegraphics[scale=0.275]{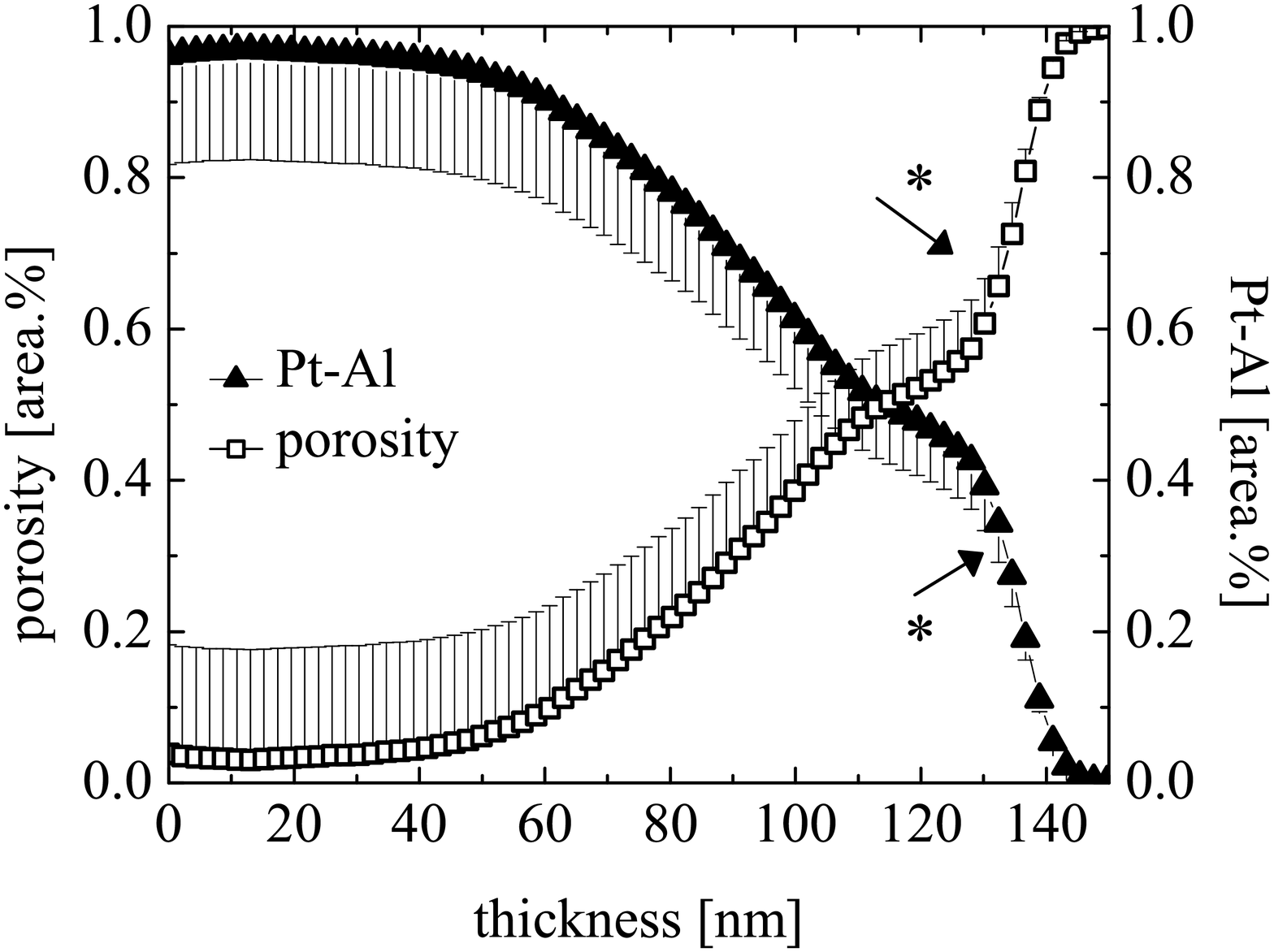}}
  \end{center}
  \caption{\ref{fig:dealloy1-a} Three-dimensional (3D) reconstruction of a dealloyed
120nm-thick Pt$_{.72}$Al$_{.28}$ film on \ce{Si3N4} obtained via FIB nanotomography \ref{fig:dealloy1-b} Developing of the mean porosity as function of the film thickness. Regions denoted by $\ast$ indicate deviations due to the dissolution of Pt which caused film shrinkage and a steeper gradient in porosity.} 
  \label{fig:dealloy1}
\end{figure}
\begin{figure}[b!]
  \begin{center}
    \subfigure[]{\label{fig:dealloy2-a}\includegraphics[scale=0.275]{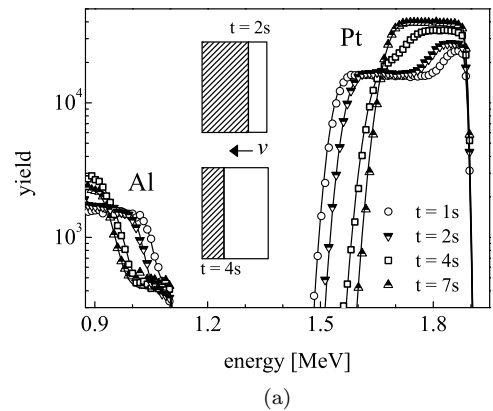}}
    \\
    \subfigure[]{\label{fig:dealloy2-b}\includegraphics[scale=0.275]{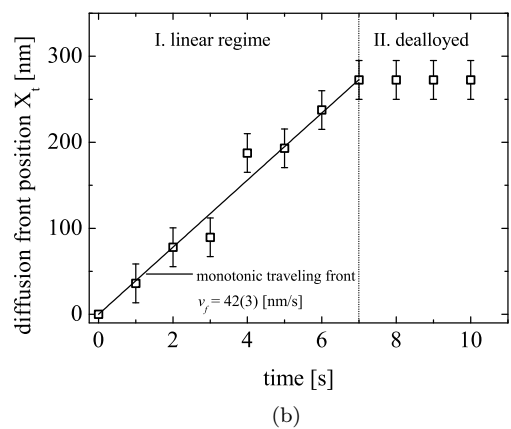}}
  \end{center}
  \caption{\ref{fig:dealloy2-a} Plot of the subtracted $2$MeV $^4$\ce{He} RBS spectra of the time-resolved dealloying of Pt/Al thin films (dots) and simulated spectra (solid lines) using RUMP. \ref{fig:dealloy2-b} Propagation front velocity $v_f=\partial_t X_t$ derived from the shrinkage-corrected front positions $X_t$ obtained by RUMP simulations} 
  \label{fig:dealloy2}
\end{figure}     
The morphological and compositional analysis of the samples was studied via FIB and RBS. Single and multiple cross-sections of the dealloyed Pt/Al systems were cut, polished and imaged using a Zeiss NVISION 40 FIB etching system.~The stacks of multiple cross-sections were aligned recursively by Stackreg~\cite{Thevenaz1} and reconstructed using AVS Express (Advanced Visual Systems Inc.). The voxel-size of the resulting tomographic images is not cubic and $2.17\times2.17\times6.5$~nm$^3$ in size. 
The compositional analysis was performed by RBS experiments using a $2$MeV $^4$\ce{He} beam and a standard silicon surface barrier detector at $165^\circ$. The background was subtracted using a common fitting procedure~\cite{Dobeli1}. The elemental composition and diffusion profiles were obtained using the RUMP program~\cite{Rump1}. For the investigated films RBS provides a quantitative one dimensional depth profile of the composition with a depth resolution of $10$ to $20$~nm~\cite{Wang1}.\\
In Figure~\ref{fig:dealloy1-a}, the spatial arrangement of the PtAl-phase as obtained by the FIB-nanotomography reconstruction after $10$~s of dealloying is shown. The dealloying results in a fine branch-like pattern with a median branch thickness of $15-20$~nm. The formed pattern is non-uniform and characterized by a strong directive gradient porosity as function of the film thickness $h$. This gradient porosity, as shown in Figure~\ref{fig:dealloy1-b}, is substantiated by calculating the porosity from multiple cross-sections of the dealloyed Pt/Al film. The porosity is increasing with increasing film thickness $h$ and in the vicinity of the film/ambient interface a zone of $\approx 25$~nm thickness is observed where the porosity evolution changes its functional shape significantly. The zone denoted by $\ast$ in Figure~\ref{fig:dealloy1-b} is assumed to originate due to the prolonged exposure to the alkaline solution that caused the nearly complete dissolution of the Al in the film which consequently led to the dissolution of the Pt and thus to a shrinkage of the film.\\
In order to further evaluate the dynamics of pattern formation, time-resolved RBS spectra have been measured and analyzed using the RUMP software. From the simulated spectra, critical time-dependent measurands like the spatial composition $\eta$, the thickness of each layer $h$, the diffusion coefficient $D$ and additional loss of Al in the dealloyed layer have been determined. The position of the dissolution front $X_t$ is calculated with respect to the initial film thickness $h_0$ and thus consideres a possible layer shrinkage in the dealloyed layer. The resulting RBS spectra and their corresponding simulation using $\eta,h,X_t$ as fitting parameters are shown in Figure~\ref{fig:dealloy2-a}.\\
The RBS spectra verify not only the dissolution of Al as function of time but also a step-like profile in the Pt-peak which can be attributed to the position $X_t$ of the traveling diffusion front during dealloying. In addition a shrinkage of the Pt/Al film in the order of $-25(3)$~nm/s is indicated, which is a direct result of the measured additional Al-loss of $0.02$~at.\%/s in the already dealloyed layer. These results are in accordance to the microstructural findings in Figure~\ref{fig:dealloy1} and present clear evidence that the formed pattern is not solely defined by the reactive processes close to the propagation diffusion front but also by diffusion processes in the bulk of the alkaline solution. As shown in Figure~\ref{fig:dealloy2-b}, the shrinkage-corrected position $X_t$ of the travelling diffusion front scales linearly with time $t$ and the corresponding velocity of this monotonic travelling front is $v_f=\partial_t X_m=42(3)$~nm/s. All relevant data have been summarized in Table~\ref{tab:table1}. \\ 
\begin{table} [t]
\caption{\label{tab:table1} Measured initial and final alloy compositions $\eta_{0,\text{end}}$, diffusion front velocity $v_{\text{f}}$, film shrinkage and the diffusion coefficient $D$.}
\begin{ruledtabular}
\begin{tabular}{ccccc}
$\eta_{0}~[\text{at.}\%]$& $\eta_{\text{end}}~[\text{at.}\%]$ & $v_{\text{f}}$~[nm/s] & $\partial_{t}h$~[nm/s] & $D$~[m$^2$/s]\\
\hline
Pt$_{.24}$Al$_{.76}$& Pt$_{.72}$Al$_{.28}$ & $42(3)$ & $-25(3)$ &$ 4.2(13)\cdot10^{-17} $\\
\end{tabular}
\end{ruledtabular}
\end{table}
The experimental observation of an initially flat liquid/film interface that evolves with time to a propagating diffusion front with a constant front velocity $v_f$ are specific characteristics of the Fisher-Kolmogorov-Petrovskii-Piskounov (FKPP) equation obeying a traveling wave solution with $u(x,t)=\psi(x-v_f\:t)$~\cite{Saarloos1}. The FKPP equation reads as follows
\begin{equation}
	\frac{\partial}{\partial t}u(x,t)=D\Delta_{x}u(x,t)+\mu u(x,t)(1-u(x,t)),
	\label{eq:fkpp1}
\end{equation}
where $D$ is a diffusion coefficient and $\mu$ represents the reaction rate of the system. The FKPP equation is a well known and widely applied nonlinear reaction-diffusion equation~\cite{Lemarchand1,Benguria1}. 
\begin{figure}[h!]
	\centering
		\includegraphics[scale=0.275]{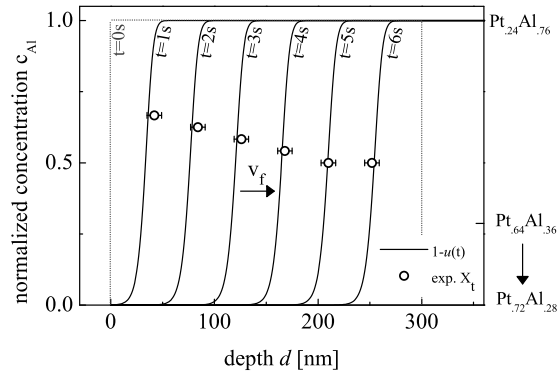}
	\caption{ Comparison of the diffusion fronts obtained by numerical solution of \ref{eq:fkpp1} in the case of $\mu=11.0(8)$~at.\%/s and $D=4.2(13)\cdot10^{-17}$~m$^2$/s with front positions $X_t$ calculated using $v_f$.}
	\label{fig:dealloy3}
\end{figure}
Using structural stability arguments~\cite{Paquette1}, it can be shown that the front velocity $v_f$ is solely defined by the reaction rate $\mu$ and the diffusion coefficient $D$ and reads
\begin{equation}
	v_{f}=2\sqrt{\mu D}.
	\label{eq:fkpp2}
\end{equation}
Using the experimental values for $v_{f}$ and $D$, the reaction rate of the dealloying system is determined to $\mu=11.0(8)$~at.\%/s. Thereafter Equation~\ref{eq:fkpp1} has been solved numerically using Mathematica with a Heaviside step function $H(x)$ as initial condition $u(0,t)=1-H(x)$. To match the experimental conditions though, the corresponding solutions is established to be $c(x,t)=1-u(x,t)$. In Figure~\ref{fig:dealloy3} $c(x,t)$ with $\mu=11.0(8)$~at.\%/s and $D=4.2(13)\cdot10^{-17}$~m$^2$/s have been plotted in comparison to the experimentally determinded front positions $X_t$. The chosen reactive diffusion equation fits nicely to the measured diffusion front positions $X_t$, although it does not feature the measured additional Al-loss in the early stage of dealloying. By choosing $c(x,t)=1-u(x,t)$ the reaction term of the RD system becomes $R(C)=\mu\:c(x,t)(c(x,t)-1)$, this kind of reaction-diffusion systems describe branching random walks~\cite{Brunet2}. In the present case the branching event is assumed to be the dissolution of Al out of the Pt/Al alloy whose space is than occupied by the alkaline solution. Thereby the spatial occupation grows linearly with time~\cite{McKean1}, as also proven experimentally in Figure~\ref{fig:dealloy3}.\\
In essence, it has been shown that consistent with the experimental findings, the dynamics of dealloying can be treated as a reaction-diffusion system. Thereby the used FKPP equation reproduces the experimental characteristics in a satisfactory manner. The RD-system is fully defined by the measured propagation front velocity $v_f=42(3)$~nm/s and the diffusion coefficient $D=4.2(13)\cdot10^{-17}$~m$^2$/s.\\ In addition, experimental evidence has been found that the pattern formation related to the dealloying process can be regarded as a superposition of the reaction-diffusion system confined to the solid/liquid interface and an additional slower dissolution process with $0.02$~at.\%/s in the bulk of the alkaline solution. This process is much slower than the propagating diffusion front but it has a severe impact on the end-form of the dealloyed pattern and the porosity distribution.\\
In an upcoming work, the obtained dealloyed films are analyzed in matters of their catalytical properties and stability during an oxygen reduction reaction (ORR).
\\
The authors greatfully acknowledge the financial supported by the Swiss Bundesamt f\"ur Energie (BfE), Swiss Electric Research (SER), the Competence Center of Energy and Mobility (CCEM) and the Swiss National Foundation (SNF). We would like to thank the EMEZ (Electron Microscopy Center, ETH Zurich) for their support. Henning Galinski would like to thank Anna Evans, Iwan Schenker and Barbara Scherrer for fruitful and stimulating discussions.

\newpage 
\bibliographystyle{apsrev4-1}
\bibliography{dealloying_arxiv}

\end{document}